\documentclass{WileyChemistry-template}

\title{\raggedright The role of surface material properties on the behavior of ionic liquids in nanoconfinement: A critical review and perspective}

\author{
\begin{minipage}{\textwidth}
	Irina Nesterova\textsuperscript{[a]}, Nikolay Kondratyuk\textsuperscript{[b]}, Yury A. Budkov\textsuperscript{[c]}, Kirill M. Gerke\textsuperscript{[d]} and Aleksey Khlyupin*\textsuperscript{[a]}
\end{minipage}
}

\newcommand{\affiliation}{
\begin{itemize}		    

\item[{[a]}] Irina Nesterova and Dr. Aleksey Khlyupin*\\
Moscow Center for Advanced Studies, 123592, Moscow, Russia\\
E-mail: irina.nesterova@phystech.su\\
khlyupin@phystech.su

\item[{[b]}] Dr. Nikolay Kondratyuk\\
Moscow Center for Advanced Studies, 123592, Moscow, Russia\\
HSE University, 101000, Moscow, Russia\\
E-mail: kondratyuk@phystech.su

\item[{[c]}] Prof. Yury A. Budkov\\
Laboratory of Multiscale Modeling of Molecular Systems, G.A. Krestov Institute of Solution Chemistry of the Russian Academy of Sciences, 153045, Ivanovo, Russia\\
Laboratory of Computational Physics, HSE University, 123458, Moscow, Russian Federation\\
E-mail: ybudkov@hse.ru

\item[{[d]}] Dr. Kirill M. Gerke\\
Moscow Center for Advanced Studies, 123592, Moscow, Russia\\
Schmidt Institute of Physics of the Earth of Russian Academy of Sciences, 123242, Moscow, Russia\\
E-mail: kg@ifz.ru

\end{itemize}
}

\renewcommand{\abstract}{Room temperature ionic liquids show great promise as electrolytes in various technological applications, such as energy storage or electrotunable lubrication. These applications are particularly intriguing due to the specific behavior of ionic liquids in nanoconfinement. While previous research has been focused on optimizing the required characteristics through the selection of electrolyte properties, the contribution of confining material properties in these systems has been largely overlooked. In this Review, we provide constructive analysis of recent developments related to the description of surface material properties impact on the ionic liquid behavior in the confinement and propose potential ways for further investigations in this direction. Although the presented advances reveal the importance of surface material properties in the application processes with confined ionic liquids, there are still a lot of issues that should be thoroughly investigated in future. We believe that this review will significantly contribute to the development of new approaches with material properties consideration for confined ionic liquid research.
}

\newcommand{\keywords}{
	Ionic liquid \textbullet\ 
	Electric double layer \textbullet\ 
	Electrode properties \textbullet\ 
	Electrode-electrolyte interface \textbullet\ 
        Nanoconfinement
}

\begin{document}

\twocolumn[\vspace{-1.5cm}\maketitle\vspace{-1cm}
	\textit{\dedication}\vspace{0.4cm}]
\small{\begin{shaded}
		\noindent\abstract
	\end{shaded}
}


\begin{figure} [!b]
\begin{minipage}[t]{\columnwidth}{\rule{\columnwidth}{1pt}\footnotesize{\textsf{\affiliation}}}\end{minipage}
\end{figure}




\section{Introduction}
\label{introduction}

The liquid phase electrolyte confined within inhomogeneously structured surfaces exhibits remarkably specific complex behavior. \cite{wang2020electrode, kondrat2023theory} It is enriched by different physical phenomena occurring within such systems, including overscreening and overcrowding,\cite{bazant2011double} underscreening,\cite{goodwin2017underscreening} structural ion reorganization,\cite{voroshylova2021ionic} anomalous capacitance,\cite{chmiola2006anomalous} nonmonotonic friction properties driven by high electric charge and normal load,\cite{di2019electrotunable} and comprehensive ion dynamics with several charging mechanisms and pore blockage at high voltages.\cite{breitsprecher2018charge} The resulting complex behavior proceeds from the combination of ion-ion and ion-wall interactions comprising strong electrostatic forces, steric restrictions, mechanical deformation, chemical aspects, and the specificity of confinement effects. 

\begin{figure*}
\centering
\includegraphics[width=1\linewidth]{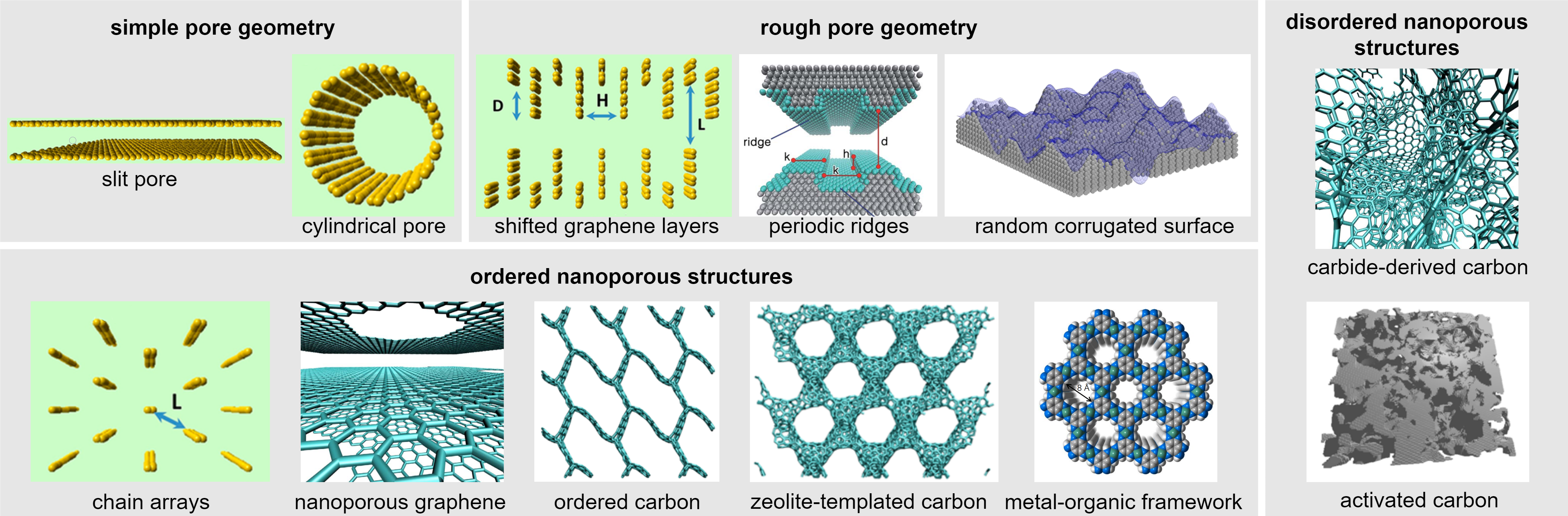}
\caption{The diversity of morphological structures of confining materials from simple and rough pore geometry to ordered and disordered carbon structures. Slit pore, cylindrical pore, shifted graphene layers, and chain arrays are reprinted with permission from Ref.\cite{vatamanu2015non}. Copyright 2015 American Chemical Society. The periodic ridges rough pore reproduced from Ref.\cite{david2017electrotunable} with permission from the Royal Society of Chemistry. The random corrugated surface is reprinted from Ref.\cite{aslyamov2017density}, with the
permission of AIP Publishing. The nanoporous graphene and carbide-derived carbon are reprinted from Ref.\cite{mendez2019performance}, Copyright (2019), with permission from Elsevier. The ordered carbon is taken from Ref.\cite{lahrar2019ionic}. The ZTC is reprinted with permission from Ref.\cite{lahrar2021simulations}. Copyright 2021 American Chemical Society. MOF picture is from Ref.\cite{feng2018robust}. The activated carbon is reprinted with permission from Ref.\cite{vasilyev2019connections}. Copyright 2019 American Chemical Society}
\label{fig:morphologies}
\end{figure*}

The variety of recent technological applications is based on the confined ionic liquid (IL) properties and behavior, including highly effective energy storage tools,\cite{wang2020electrode} electrotunable lubrication,\cite{palacio2010review} nanoparticle dispersion,\cite{perkin2012ionic} and catalysis processes.\cite{zhang2017nanoconfined} That induces extensive research aimed to describe and predict the properties of such systems for application performance optimizations during past decades.\cite{ fedorov2014ionic, vatamanu2017charge, hartel2017structure, jeanmairet2022microscopic} However, the main focus was previously paid to the properties of electrolytes, such as ion size or valency asymmetry, presence of solvent, formation of ion pairs, and the others, considered to be confined within flat
or smooth surfaces. Real materials, which confine ionic liquids, usually have complex inhomogeneous surfaces. Actually, atomic-force microscopy (AFM) experiments are required to study the structure of non-flat surfaces, \cite{migliorati2022crystalline} and highly efficient modern electrodes, like carbide-derived carbons (CDC), have disordered multi-scale porous structures with local structural features.\cite{merlet2012molecular} The electrode surface is usually considered only as a boundary condition in the analytical and theoretical works,\cite{bazant2011double, henderson2011density, budkov2021electric} i.e., implying an ideally polarizable metal surface, neglecting its real chemical structure and the ability to respond to the electric field. Additionally, only a few works considered electrode deformation during charging, \cite{xu2020effects} although confined liquid can produce huge disjoining pressures.\cite{gurina2022disjoining} Therefore, not enough attention has been paid to uncovering the contribution of substrate properties to the behavior of confined ionic liquids.

Moreover, when considering confined electrolytes, it is crucial to distinguish the reasons for observed effects: whether they arise from material characteristics or confinement conditions. Despite this, it is well known that electrode properties can dramatically change the properties of confined ionic liquids, such as differential capacitance\cite{xing2012nanopatterning, bacon2023key} or friction coefficient. \cite{mendonca2013nonequilibrium, david2017electrotunable} Properly describing the contribution of material properties represents a significant gap in modern electrochemistry. However, an accurate description of substrate properties is mainly available only through molecular dynamics (MD) simulations, even though there are also particular challenges. Thus, the development of new theoretical and analytical approaches that reflect the fundamental impact of material properties is highly desirable.

In this review, we collect the latest cutting-edge works, which reveal the contribution of substrate material properties, focusing on the surface morphology and multi-scale nanoporous geometry, consideration of mechanical material properties, and chemical structure of confining material. We provide our vision on weakly covered issues, thoroughly carry out an analysis of current observations, raise a lot of questions, and give potential strategies for future developments in this field. The Review has the following structure, first, we discuss in detail how the geometrical structure of confining material impacts ionic liquid properties and provide a detailed analysis of the literature background. Then, we move to account for the mechanical properties of the electrolyte confined between two solid substrates and give the description of theoretical approaches applied to this problem. Next, we give an overview of recent observations related to various chemical properties of confining materials and their contribution to the confined electrolyte behavior. Finally, we provide an extensive vision for the existing challenges and possible extensions, binding them with particular, analytical, theoretical, and numerical approaches, that should be developed to advance the understanding of confined electrolyte behavior.

\section{Material surface morphology and multi-scale nanoporous geometry}
\label{roughness}

\begin{figure*}
\centering
\includegraphics[width=0.95\linewidth]{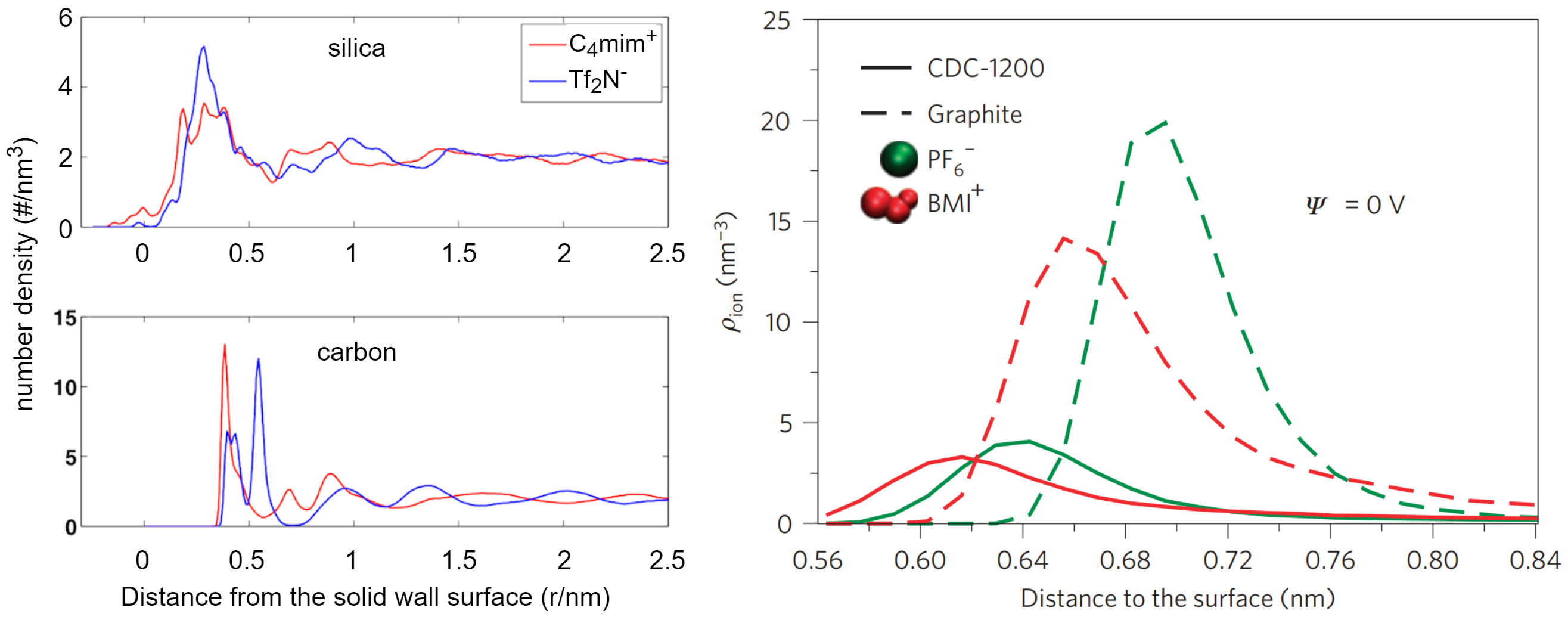}
\caption{The impact of surface morphology on IL structure. On the left, the structure of C$_4$mim$^+$ Tf$_2$N$^-$ ionic liquid is presented near smooth carbon and slightly corrugated silica surfaces, adapted with permission
from Ref.\cite{li2013dynamic}. Copyright 2013 American Chemical Society. On the right, the structure of BMI$^+$ PF$_6^-$ ionic liquid is presented near the planar graphite electrode and CDC-1200 \cite{merlet2012molecular}}. 
\label{fig:structure}
\end{figure*}

\subsection{Morphological diversity}

To set the scene, we start with a brief literature review of works, where IL are studied in nanoconfinement with various surface geometry and nanoporous structures. Along with that, we analyze the possible extensions, which will provide more comprehensive understanding for the observed results.

There are several works on the ionic liquid confined with rough pore walls reflecting real disorder atomistic pore wall roughness for slit \cite{ori2014ionic} and cylindrical \cite{li2013dynamic} silica nanopores. In contrast to smooth graphene surfaces or carbon nanotubes, silica surfaces usually have nanoscale corrugation. However, the scale of surface structural inhomogeneities is too small $\sim0.1$ nm, so they only shift, broaden and decrease ion density peak near the pore wall compared to flat one. To understand the impact of the electrode surface structure, it would be better to carry a range of experiments with more significant scales of surface structural inhomogeneities of the order of the size of ion and pore widths. This would allow to evaluate its effect on electric double layer (EDL) structure. 

Later, the ionic liquid was considered in the confinement with more pronounced surface corrugation. \cite{vatamanu2015non, david2017electrotunable} For instance, inside rough pores with widths between 8 and 20 Å made of two prismatic graphite surfaces having grooves with depth of 5.1 Å and distance between grooves 7.3 Å.\cite{vatamanu2015non} Within this system, rough edges trigger better ion separation, change character of ion transport mechanism, and provide higher integrate capacitance. Besides, rough confinement was presented as two slabs with periodic rectangular ridges having height of 1 nm and a distance between them of 2 nm. The distance between slabs varied from 2 to 6 ion layers, i.e., about the order of surface corrugation. In such a system, the increase of the friction coefficient between corrugated plates as well as its dependence on shear direction was observed. \cite{david2017electrotunable} These works provide important insights on the impact of electrode surface roughness on the behavior of ionic liquid in the confinement along with their properties such as capacitance or viscosity. Meanwhile, the considered structures have too simple character consisted of rectangular grooves or shifted sheets of graphene. However, the roughness of real surfaces has more complex forms and random character of localization. 

The most realistic and significant rough confinement was presented between rough metal (iron) surfaces having a distance from 30 to 60 Å from each other and containing local structural features with non-ideal geometry of truncated cones having a radius 12 Å and 10 Å, and a height of 8 Å. \cite{mendonca2013nonequilibrium}  As a result, the electrode surface structural inhomogeneities lead to a higher friction coefficient due to the promotion of ion reorientation with higher charge ordering. Therefore, that work makes a significant contribution, although until now there are too many questions about the impact of structural features on the properties of ionic liquid in the confinement. Particularly, how the form, size, and number of surface structural features modifies IL behaviour. We hope that our concept will trigger extensive research on the investigation of rough confinement with various scales, structures, and geometries of electrode surfaces.

In order to study nanoporous electrodes with an ordered structure, researchers have considered the following materials: carbon chains, \cite{vatamanu2013increasing} nanoporous perforated graphene, \cite{mendez2019performance} metal-organic frameworks (MOF), \cite{bi2020molecular} and zeolite-templated carbon (ZTC). \cite{liu2019carbons} 
Carbon chains are shown to provide the highest integral capacitance for systems without confinement. \cite{vatamanu2013increasing} Their performance in the confinement conditions was also compared to slit pores, while there is no systematic comparison for various electrode structures. Particularly, carbon chains have a really interesting structure due to the relatively high role of steric repulsion while the role of ion-wall attraction remains relatively small. However, it definitely requires the inclusion of mechanical properties, because such structure is barely maintained during the charge storage process. The nanoporous perforated graphene structure consists of a graphene sheets array with randomly located holes in the sheets.\cite{mendez2019performance} This structure concedes CDC electrodes, cause it limits the freedom for ion localization because of its uniform structure. Such a well-defined structure is useful for the investigation of pore percolation and pore size distribution impact on charging dynamics due to the simplicity of structural modifications such as varying size, number, and location of the holes on graphene sheets and the distance between them. 

MOFs are great electrodes for the validation of numerical modeling with experiments due to the monodisperse character of porous space and controllable structures. They were applied to analyze the capacitance characteristics of electrodes and showed DC form transition with the increase of pore size. \cite{bi2020molecular} In contrast to MOFs, ZTC have complex ordered structures with pore size distribution and well-defined pore geometry. It allows to investigate the influence of electrode structure on charging performance. For example, charging dynamics become worse with pore limiting diameter, while no correlation of capacitance with average pore size was found.\cite{liu2019carbons} 

The CDC-based electrodes have a highly disordered and amorphous structures with local structural features and consist of pores with different scales. \cite{palmer2010modeling} The behavior of ionic liquid in such complex systems has a non-trivial character, as it is driven by various physical phenomena including electrostatic, dynamic, and confinement effects. The investigation of ionic liquid behavior in CDC microporous structure is actively studied by MD simulations.\cite{merlet2012molecular, merlet2013highly, pean2015confinement, mendez2019performance}
The disordered character of CDC structure allows coions to be placed far from the surface leading to higher capacitance. \cite{mendez2019performance} Besides, it modifies the structure of the ionic liquid, preventing the occurrence of overscreening effect, which, in turn, leads to capacitance increase together with local structural features, like graphite domains.\cite{merlet2012molecular} Another study showed that local morphological characteristics define the level of desolvation and local stored charge. \cite{merlet2013highly} In that work, local morphological characteristics were described with a degree of confinement, which is a fraction of the angle around the ion occupied with carbon atoms. 
Additionally, it was shown that the process of ion desolvation in the nanoconfinement is found to be the fastest, compared to ion diffusion and adsorption, \cite{pean2015confinement} while it should depend on the interaction forces between the ion, solvent, and pore walls. 

Summing up, there are two main mechanisms in the nano-\\porous electrode to improve charge storage performance. The first one is related to the formation of a new optimal structure for ionic liquid created with electrode surface morphology and steric restrictions. It is still unclear whether it will always be more efficient and how to control it? The second mechanism relies on the confinement effect, which drives ion desolvation, but it requires more accurate fundamental study by theory describing local structural features of nanoporous electrodes.

\subsection{Confined ionic liquid specificity}

Now, we acknowledge that morphological characteristics of charged surfaces, which confine ionic liquids, contribute to IL structure and properties in the nanoconfinement. Therefore, the material surface structure and the geometry of nanoporous space become a key to control and optimize the characteristics of the confined IL for particular applications. As for the investigations of ionic liquid in the confinement with inhomogeneously structured electrodes, recent developments and observations obtained by molecular dynamics simulations are well ahead the results of analytics and theory.\cite{kondrat2023theory} In particular, researchers are mainly focused on nanoporous materials with complex structure, rather than rough confinement, motivating it with representation of real electrode materials with high surface area, such as CDC.\cite{merlet2012molecular} Those studies capture the whole information about electrode structure, such as pore connectivity, pore size distribution, surface area, local morphological features, and give their total contribution into the electrode-electrolyte system properties. Nonetheless, to connect theory with simulations and to distinguish effects from different morphological characteristics of electrode, it would be better to "step back" and focus on rough confinement or nanoporous materials with simple or slightly disordered structures, rather than nanoporous systems. Along with that, researchers should extensively develop theoretical and analytical approaches to boost the design of optimal supercapacitors and IL lubricants.

Let us clarify how the surface structure and the geometry of porous space change the behavior, structure, and properties of ionic liquid in the confinement. In short, the pattern of the material surface changes the ion's organization near its surface, i.e., allowing some ions to come closer in the surface structural holes, pushing other ions away by surface structural ridges, and also it can tend ions for a particular angular orientation optimal for the system free energy minimization.\cite{vatamanu2012molecular, xing2012nanopatterning} So, the modified structure of IL could provide a better charge accumulation by improved ion separation near the electrode surface\cite{vatamanu2012molecular, xing2012nanopatterning} or accelerate ion transport by disordered character of confined ionic layers structure.\cite{li2013dynamic} Therefore, the structure of IL define all the required properties for technological applications, which, in turn, is restricted by material surface geometry.

In the ultra-tight confinement, so small that ions do no enter the pore, the ionophobic conditions are created, which provide higher energy densities. \cite{lian2016can, kondrat2023theory} Thus, the electrode geometry and surface structure allow for creating ionophobic electrodes with higher energy storage performance.\cite{vatamanu2013increasing} Notably, the idea of creating ionophobic conditions by surface structure refers to the wetting phenomenon, where nanoscale surface corrugation is widely applied for the surface protection from liquid devastation impact. In turn, ionophobicity is the electrostatic analogue of wetting.\cite{berim2008nanodrop, wang2015recent} It was shown both theoretically\cite{khlyupin2023molecular} and numerically \cite{vatamanu2011influence, xing2012nanopatterning} that the electrode surface roughness transforms the form of differential capacitance (DC), i.e., modifies its ionophobicity. Particularly, the transition from bell to camel reflects the growth of ionophobicity, and the transition from camel to bell occurs at more ionophilic surfaces, according to density functional theory (DFT) calculations.\cite{lian2016can} These observations prove that the morphology of electrode surface can significantly improve energy storage.

We tend to agree with the vision of J. Vatamanu \cite{vatamanu2015non} about the investigation of liquids confined with structured surfaces. The author stated that, first of all, one needs to investigate the electrode morphological impact without confinement conditions and then to consider the system in the confinement. Therefore, we need to  distinguish clearly the effect induced by confinement conditions from those driven by material structure. In what follows, we provide the review on the effects from the confinement and material structure separately and then give the vision of their total contribution. According this scheme, we analyze the impacts of confinement and morphology on energy storage and ion transport.

\subsection{Energy storage and capacitance}

\begin{figure*}
\centering
\includegraphics[width=0.85\linewidth]{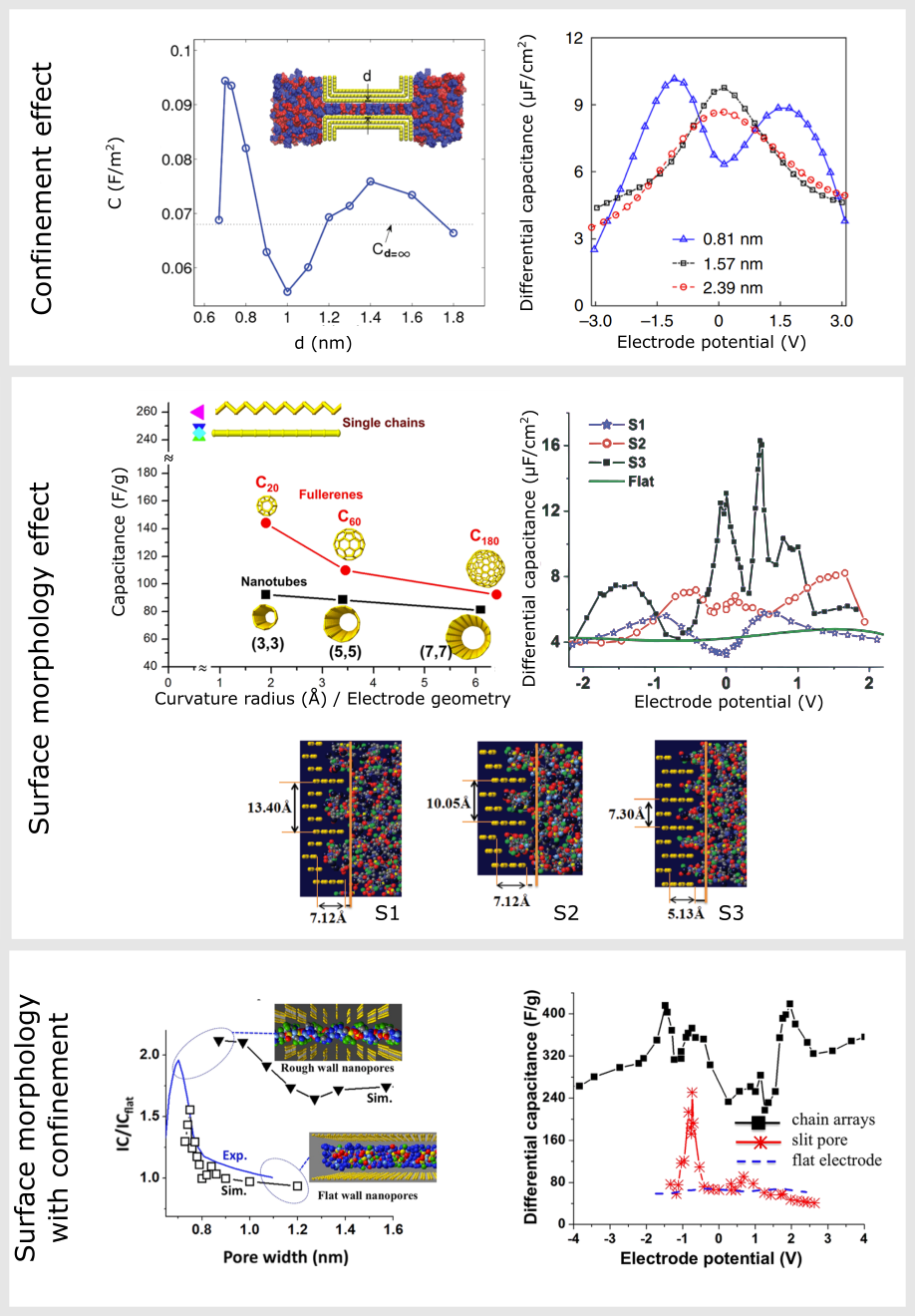}
\caption{The distinguished and combined views on the effects of confinement conditions and surface morphology on capacitance characteristics. On the top, the behavior of the integral and differential capacitance of IL depends on pore width for slit confinement condition. The relation of capacitance with pore width is reprinted with permission from \cite{feng2011supercapacitor}. Copyright 2011 American Chemical Society. The DC profiles for different pores widths are taken from Ref.\cite{bi2020molecular}. In the center, the effect of electrode surface morphology on the capacitance characteristics is presented. The variation of capacitance on the surface geometry is adapted with permission from \cite{vatamanu2013increasing}. Copyright 2013 American Chemical Society. The DC variation near structured electrodes with different structures shown below are reprinted with permission from \cite{xing2012nanopatterning}. Copyright 2012 American Chemical Society. On the bottom, the behavior of capacitance characteristics within structured confinement. The capacitance enhancement in the rough nanopore is reprinted with permission from \cite{vatamanu2015non}. Copyright 2015 American Chemical Society. The differential capacitance profiles near the flat electrode, in the slit confinement, and in the chain arrays electrode are reprinted with permission from \cite{vatamanu2013increasing}. Copyright 2013 American Chemical Society.}
\label{fig:capacitance}
\end{figure*}

In this part, we focus on the electrode structure impact on the capacitance of ionic liquid in the confinement conditions. As it has been already mentioned, it is important to distinguish the confinement effects on capacitance from the effect of electrode structure. In this regard, we first give a current vision of confinement impact on capacitance within simple geometries of electrodes. Then, we review the electrode structure influence on capacitance without confinement. After that we summarize the studies that takes into account both the electrode structure and the confinement.  Finally, we give some possible extensions for the analytics and theory.


The description of ionic liquid behavior and properties in flat confinement is well-developed both theoretically and numerically. For more information on this issue we kingly ask the readers to address a great review on theory and simulation of ionic liquid in the confinement by S.Kondrat et al.\cite{kondrat2010superionic} Briefly, in the confinement ions are subjected to the superionic state that is characterized by the redaction of interionic interaction, ion desolvation, and these phenomena explained observation of anomalous capacitance in very tight pores comparable with the size of the ions.\cite{chmiola2006anomalous}
This anomalous behavior corresponds to a sharp capacitance growth with a decrease of pore size, while in the traditional view it should reduce.\cite{chmiola2006anomalous} This behavior of capacitance can be reproduced analytically by mean-field model, \cite{kondrat2010superionic} theoretically by classical density functional theory, \cite{jiang2011oscillation} and by Monte-Carlo (MC), \cite{kondrat2011superionic} and molecular dynamics  simulations. \cite{feng2011supercapacitor} Interestingly, mean-field model tends to overestimate experimental capacitance, while DFT, MC, and MD results, in contrast, underestimate them. 

Later, the dependence of capacitance on pore width in slit confinement was shown to be more complex and has an oscillatory character.\cite{jiang2011oscillation, feng2011supercapacitor, wu2011complex} If we consider pore widths smaller than the ion size then the capacitance drops, because of the steric restriction for ion to enter the pore. It should be noted that the same DC behavior was observed for polyelectrolyte solutions and polymerized ionic liquids with long polymer chains in charged slit-like pores.~\cite{budkov2023macroscopic,budkov2023dielectric} For the pore widths larger than solvated ion, the oscillatory behavior of capacitance was observed. It is explained by interference of EDLs on the pore walls or modification of the charging mechanism. If we consider corrugated pore walls that modify the structure of ionic liquid near the electrode surface, then capacitance dependence on pore width may lose its oscillatory behavior. This hypothesis seems to be worth checking both theoretically and numerically. 

As for the differential capacitance of ionic liquid in the confinement condition, it was described analytically only for simple ionic liquid structures, such as a 1D system within ultra-tight cylindrical pores, where one row of ions can locate, or a 2D system inside ultra-tight slit pores with 1 layer surface of ions. \cite{kondrat2023theory} The Ising model\cite{kornyshev2013simplest} showed a camel shape for DC of the 1D system and no effect on DC shape with pore width increase. Later, DFT calculations demonstrate DC form transition from camel to bell shape with pore width increase for 1D system \cite{kong2015density} and DC growth for a higher degree of confinement in the pore, i.e., slit < cylindrical < spherical.\cite{ma2014classical} However, these DFT works didn't consider the superionic state of ionic liquid in the confinement. Besides, it will be interesting to carry a more detailed analysis on DC transformation with the pore widths increase together with corresponding analysis of ionic liquid structure modification in the confinement. For example, when IL has the structural transition from one layer to two layers, how it is reflected in DC? This was explored in the MD study, where it was revealed that when IL undergoes the transition from a multilayered structure to one layer, the cumulative ion density becomes nonlinear and leads to a high DC variation.\cite{xing2013atomistic} Similar results were shown by Monte Carlo simulations, i.e., flattening of DC-potential curve with the increase of the pore width, which is probably related to modification of the charging mechanism. \cite{kondrat2011superionic} 

Another important issue related to capacitance in the confinement is the interpretation of capacitance behavior with the charging mechanism. Mean-field model \cite{kondrat2010superionic} shows that at low voltages differential capacitance is constant and charging proceeds due to ion swapping, keeping the cumulative ion density constant, while coion density slowly decreases. Then, DC sharply reduces because of phase transition from coion-rich to coion-deficient phases, following by the extensive coion desorption from the pore leading to the drops of both cumulative and coion density. After that, during charging the cumulative ion density slowly grows, reflecting counterion adsorption that leads to a slow decrease of DC.   More recent MD study \cite{wu2012voltage} shows similar behavior for integral capacitance with ion swapping at low voltages, where the maximum capacitance related to coion desorption and followed by the capacitance decrease corresponding to the counterion adsorption. However, the MC study \cite{kondrat2011superionic} demonstrates that the DC peak corresponds to the maximum total ion density in the pore and the following charging occurs due to the ions swapping that do not coincide with the observations mentioned above. Despite there are several works that consider this issue analytically and numerically, their observations seems not to agree with each other,\cite{kondrat2010superionic, kondrat2011superionic, wu2012voltage} so it requires further investigations. 


The effect of the electrode surface structure on the capacitance without confinement is mainly studied with MD simulation, although there are several analytical models that can describe such systems. 
The first attempt to analytically describe EDL capacitance on rough surfaces was made by solving the Poisson-Boltzmann equation with an incorporated roughness function, reflecting the scale of surface corrugation, as the scale factor for the capacitance. \cite{daikhin1996double, daikhin1998nonlinear} 
As a result, they obtain higher capacitance near the rough electrode. However, the separate consideration of compact and diffusive layers are not applicable in the case of dense ionic liquids and cannot accurately describe the phenomena occurring at the molecular scale, such as ion separation. Another attempt to describe EDL near rough surfaces analytically consists of two main steps. The first step stands for the derivation of modified expression for ion density, including steric restrictions from the electrode surface. \cite{aslyamov2021electrolyte} Second, the analytical solution for the Modified Poisson-Fermi equation with an application of this expression is built with several limiting cases: (i) where ion separation occurs mainly due to ion size asymmetry and surface roughness are insignificant, (ii) ion separation is induced by the strong surface roughness and the presence of ion size asymmetry. \cite{khlyupin2023molecular} This model successfully reproduced several essential MD observations, such as DC nonlinear behavior with temperature, DC variation with ion size asymmetry, DC transition from camel to bell near the rough surface, and even the formation of additional DC peak induced by ion reorientation.

The effect of the electrode surface roughness on ionic liquid structure and capacitance has been widely studied by MD simulations. \cite{vatamanu2011influence, vatamanu2012molecular, xing2012nanopatterning, vatamanu2013increasing, hu2013molecular, hu2014comparative} It was proved that the electrode structure drives ion rearrangement providing a prominent way to control and improve the charge storage performance. In all these studies, structured electrodes showed higher DC and larger DC variation with potential in comparison with flat electrodes.\cite{vatamanu2011influence, xing2012nanopatterning, vatamanu2012molecular, hu2013molecular, hu2014comparative} This effect becomes more prominent for more asymmetric ionic liquid \cite{hu2013molecular} and is explained by faster counterion accumulation and ion segregation.\cite{vatamanu2012molecular}  Additionally, the structure of the electrode can transform DC-potential curve from camel to bell shape\cite{vatamanu2011influence, hu2014comparative} or in another complex way, \cite{xing2012nanopatterning} but it requires proper relation of ion size to the scale of surface roughness. Actually, the pattern of electrode surface should be similar to the ion size in order to provide optimal capacitance characteristics.\cite{xing2012nanopatterning}
Interestingly, optimal electrode structure was also observed with the single chain electrode, which provides the highest capacitance compared to slit and rough surfaces or carbon tubes and spheres.\cite{vatamanu2013increasing} As there was no extensive research on different morphology of surface roughness, this statement looks controversial. Another unknown is a persistence of observed results in the confinement conditions.

\begin{figure*}
\centering
\includegraphics[width=0.8\linewidth]{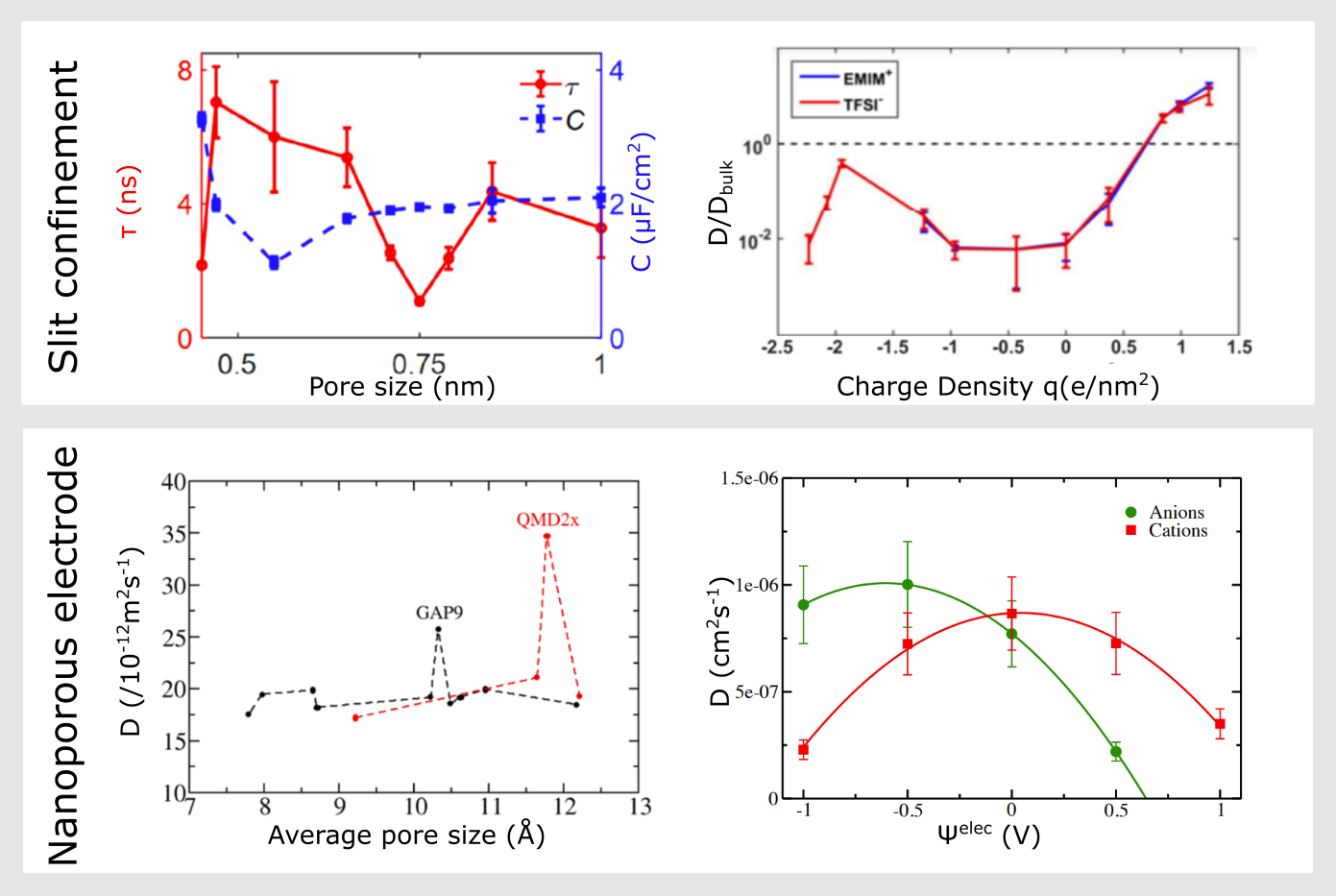}
\caption{The behavior of ion diffusion with pore size and charge on the electrode in the slit confinement and nanoporous structures. On the top, ion diffusion within slit confinement conditions. On the left, the nonmonotonic behavior of charging time with pore size (red line) is reprinted with permission from \cite{mo2020ion}. Copyright 2020 American Chemical Society. On the right, the nonmonotonic behavior of diffusion coefficient with the charge on the electrode is reprinted with permission from \cite{he2016importance}. Copyright 2016 American Chemical Society.  On the bottom, ion diffusion behavior within nanoporous electrodes is shown. On the left, the diffusion coefficient within ordered (GAPs) and disordered (QMDs) nanoporous strictures for different average pore sizes are shown \cite{lahrar2019ionic}. On the right, the diffusion coefficient variation with charge on the electrode is reproduced from Ref. \cite{pean2015confinement}}
\label{fig:diffusion}
\end{figure*}


The capacitance characteristics of nanoporous structured electrodes were studied by MD simulations for various porous structures, like rough pore, \cite{vatamanu2015non} carbon chains,\cite{vatamanu2013increasing} perforated graphene sheet, \cite{mendez2019performance} CDC, \cite{merlet2012molecular} and ZTC.\cite{liu2019carbons} For atomically rough pores, the behavior of integral capacitance with the change of nanopore width is qualitatively similar to that of the smooth slit pores.\cite{vatamanu2015non} It has oscillatory character with the broader maximum peak occurring at smaller pore widths, and DC maximum now occurs at lower potentials. Notably, the contribution of electrode structure to capacitance is found to be greater than the contribution of confinement effects. It raises questions whether the similar character of capacitance is occasional? How it would change for different surface geometries? How it quantitatively differs from the slit one? Does the optimal pore width keep the same for corrugated and slit pores, or how is it changed? Does the effect of roughness improve in the confinement compared to an open surface? These issues require a more detailed investigation.

The carbon chains showed a higher capacitance than slit pores or CDC. \cite{vatamanu2013increasing} Besides, there are optimal values for the distance between chains that ensure maximum capacitance. As it was mentioned before, the consideration of the mechanical properties the carbon chains is required to reflect real capacitance opportunities of this structure. Moreover, in our opinion, the impact of chain thickness on the confined ionic liquid capacitance as well as the consideration of the electrode material mechanical properties deserve to be investigated.

The perforated carbons showed worse surface capacitance than graphite sheets. Moreover, it was shown that the larger holes give less capacitance.\cite{mendez2019performance} Probably, it is  related to the smaller specific surface area. It seems to be true only if there is no confinement effect between graphene sheets. Consequently, perforated carbons concede CDC electrodes, because of the disordered character of CDC structure, which allows coions to be placed far from the surface leading to higher capacitance. Interestingly, would the observed result change if we vary the distance between graphene sheets similar to how it was done for carbon chains? \cite{vatamanu2013increasing} Moreover, for these systems it is worth to include the electrode mechanical properties to describe electrode expansion during charging and to study the behavior of the capacitance at these conditions. 

For ZTC electrodes, there was no clear correlation of capacitance with common structural descriptors, such as void fraction or average pore diameter.\cite{liu2019carbons} However, in our opinion, these morphology characteristics are too "average". One should consider more complex and accurate structural descriptors for such study. Meanwhile, it shown found that the structural feature called "pocket" allows counterions to be optimally located to compensate electrode charge, boosting the electrode capacitance.

For CDC electrodes, their complex disordered structure worsens ion separation and eliminates the overscreening effect, causing a boost of capacitance.\cite{merlet2012molecular} Besides, the authors compared two types of CDC with similar pore size distribution and average pore sizes, but different local features, such as small graphitic domains, which provided a capacitance increase of 43\%. Therefore, we can see how the local electrode microstructure can improve the electrode capacitance.

Other two important structural properties of nanoporous electrodes are pore connectivity and the character of pore size distribution (PSD). \cite{kondrat2012effect, vasilyev2019connections, mendez2019performance, mo2022symmetrizing, liu2024structural} It was shown that there is optimal pore width, determined as the size of a desolvated ion, which leads to higher energy density. Besides, PSD narrowing gives energy density improvement, so monodisperse porous materials are preferable for optimal energy storage.\cite{kondrat2012effect} The opposite results were observed in the MD simulations, \cite{mendez2019performance} where CDC showed better performance than equally interlayer-spaced graphene sheets, providing higher surface capacitance. It was related to a more optimal ion localization in the porous space due to the nonuniform structure of CDC. This disagreement is probably related to non-optimal interlayer distances between graphene sheets for the considered IL. Another study \cite{liu2024structural} shows no correlation between capacitance and the pore size distribution or specific surface area of nanoporous carbons for a range of nanoporous carbons. Instead, capacitance corresponds to the structural disorder and the size of graphene domains. The problem of physical processes integration from one pore to the nanoporous structure is not trivial.\cite{ma2018bridging, da2020reviewing} Previous studies demonstrated that well-percolated porous materials provide higher energy density. \cite{vasilyev2019connections} Additionally, the application of the equivalent capacitance principle was proved for the system of two asymmetric pores separated with bulk IL.\cite{mo2022symmetrizing} It is worth to compare its performance on the nanoporous electrode with ordered and disordered structures with PSD. In addition to that, theoretical works are awaited to investigate which pore first reacts to the changes in bulk conditions in the system pore+pore+bulk. The integration of pore network modeling might give some insights on the phenomena at nano and micro scales.

\subsection{Ion Diffusion and Charging Mechanism}

The dynamical properties of ionic liquid in the confinement are studied by means of analytical approaches, classical density functional theory, and MD simulations, but the main focus is still kept on the slit confinement.\cite{kondrat2023theory} We suggest that the exploration of the ionic liquid dynamical behavior in structured nanoconfinement should follow a similar scheme as in the discussion of capacitance, i.e., we need to distinguish the effect of the confinement and electrode surface geometry from one another. Although majority of works describe ion diffusion in slit confinement, the comprehensive view of ion dynamics in the nanoporous electrodes is still missing.

Researchers classify four charging mechanisms in the confinement: linear, square root, and two exponential, which occur consistently with charging time. \cite{kondrat2023theory} These regimes are dictated by the processes going inside the pore. In the linear regime, the pore becomes overcrowded with counterions, making coions break out of the pore through a dense counterions layer and the ion swapping that slows down with the charging time, which is probably related to square root and first exponential regime. The last exponential regime reflects counterion desorption during the equilibration between pore and bulk. 

Ion diffusion in the confinement is found to worsen with pore width decrease, especially in the ultra-tight pores with nearly 1-2 ion diameter, according to the mean-field model, dynamic DFT, and MD studies results. \cite{kondrat2013charging, aslyamov2022relation, gading2022impact, otero2020nanoconfined,gurina2023transport} Other works on molecular dynamics simulations show that the confinement induces a nonmonotonic dependence of ion diffusion on the pore width and the electrode charge.\cite{kondrat2014accelerating, he2016importance, mo2020ion} In the ultra-tight pore, where only one molecular layer can be located, for small and high charges on the pore wall the structure of ionic liquid inside the pore has clear order, leading to poor ion diffusion, which in two orders worse than in the bulk. However, at moderated charges, the structure of ionic liquid becomes disordered, and ion transport improves and exceeds the bulk one. \cite{kondrat2014accelerating} Such structural modification due to the changes in the ions angular orientation is also determined by the pore width.\cite{gurina2023transport} Similar conclusion was made for ion transport dependence on pore width. \cite{mo2020ion} In that work, ion transport improved when ionic liquid undergo a structural transition from one layer structure to two layers. Another work \cite{he2016importance} shows that wider pores with several molecular layers present no improvement of ion diffusion at any potential on the electrode. It is, thus, obvious that ion diffusion is mainly controlled by the structure of the ionic liquid inside the pores, which, in turn, depends on the electrode charge and the relation of the pore size to the ion diameter.

Currently, the ion dynamics description in the nanoporous electrodes were performed only by MD simulations. Perhaps, if we place ionic liquid into rough pores, where its ordered structure will be broken by electrode surface corrugation, it will improve ion transport in comparison to slit confinement. This suggestion might explain faster dynamics observed in the MD study\cite{pean2016understanding} of the CDC electrodes charging, where no overcrowding appears by counterion adsorption at higher potential difference. In common, observations similar to ones in the slit confinement, were obtained in nanoporous materials. The smaller average pore size leads to the slowest charging \cite{pean2014dynamics, liu2019carbons, lahrar2019ionic} and nonmonotonic dependence of ion diffusion on the electrode potential. \cite{pean2015confinement} 
Actually, in the MD study\cite{lahrar2019ionic}, nonmonotonic behavior of diffusion coefficient with the pore size was observed for several structures of nanoporous carbons. As for charging mechanisms, counterions are found to move to local structures with higher confinement, while coions pretend to locate new low confinement sites.\cite{pean2016understanding}
Similar to the slit confinement case, the results on the ion diffusion in nanoporous materials disagree with one another, so these systems required further investigations. Therefore, it will be interesting to theoretically describe the nonmonotonic behavior of ion diffusion observed in the MD studies.

\section{Mechanical properties of substrate material}
\label{mechanics}

\begin{figure*}
\centering
\includegraphics[width=0.85\linewidth]{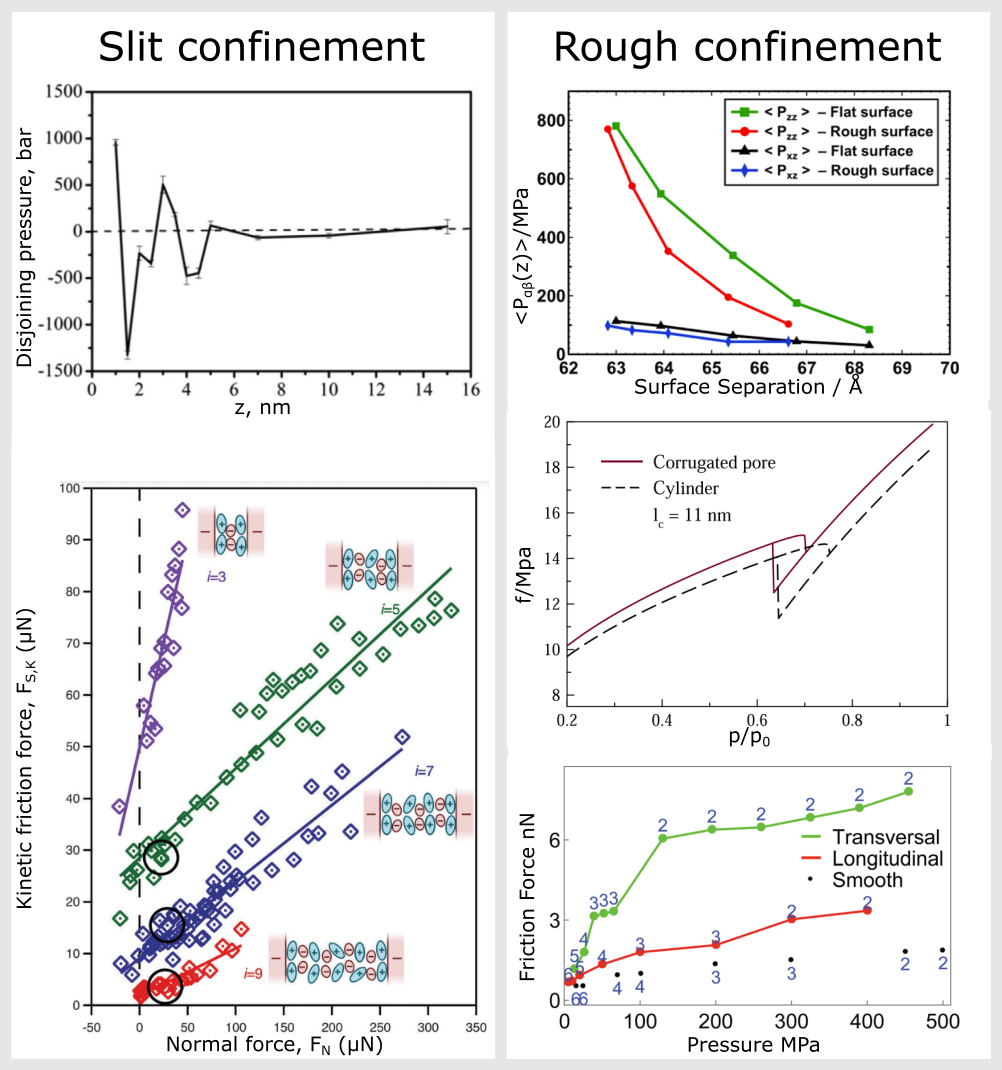}
\caption{The behavior of mechanical properties within slit and rough confinement. On the left, at the top, the disjoining pressure with slit pore width is reprinted from Ref.\cite{gurina2022disjoining} with permission from Elsevier. On the left, in the bottom, the quantized friction with normal load for slit confinement is shown from Ref.\cite{smith2013quantized}. On the right, at the top, the comparison of stress tensor components for flat and rough confinement are reprinted with permission
from Ref.\cite{mendonca2013nonequilibrium}. Copyright 2013 American Chemical Society. On the right, in the center, the comparison of disjoining pressures for cylinder and corrugated pores are reprinted from Ref.\cite{kolesnikov2020adsorption}, with the
permission of AIP Publishing. On the right, at the bottom, the deviation of friction forces depending on confining surface structure is reproduced from Ref.\cite{david2017electrotunable} with permission from the Royal Society of Chemistry.}
\label{fig:mechanics}
\end{figure*}

The mechanical properties of solid nanostructures are essential to account for an accurate description of processes with confined ionic liquids. It mainly appears in the research on the deformation of the nanoporous electrodes during the charging process~\cite{chen2022porous, koczwara2017situ} and investigation of ionic liquid thin films confined between two surfaces or a surface and the cantilever for the surface force measurements (SFM) and AFM experiments.~\cite{gebbie2013ionic, smith2016electrostatic, kumar2022absence} For such systems, it is important to describe the mechanical stress tensor, as well as the electrostatic characteristics, and consider their collective contribution to the confined ionic liquid behavior. By calculating the local stress tensor, we can determine physically significant properties such as solvation (disjoining) pressure~\cite{evans1987phase} and tangent (shear) stresses. The solvation pressure can be used to estimate the deformation of porous materials,~\cite{gor2017adsorption,kolesnikov2021models} which is crucial for batteries and supercapacitors that utilize microporous electrodes saturated with liquid-phase electrolytes.~\cite{koczwara2017situ,chen2022porous,da2020reviewing,li2018high,kolesnikov2022electrosorption} In turn, the shear stress is useful for characterization of friction properties of ionic liquid films, which are applicable in nanolubrication due to the ability to control their properties by surface charge.\cite{mendonca2013nonequilibrium, david2017electrotunable, di2019electrotunable} Additionally, the stress tensor can be used to calculate the macroscopic force exerted on charged macroscopic conductors or dielectrics immersed in liquid phase electrolytes.~\cite{kolesnikov2022electrosorption,ruixuan2023electrostatic} Furthermore, understanding the difference between normal and tangential stresses in an ionic liquid at the interface with an electrode will allow us to calculate surface energy,~\cite{derjaguin1987derjaguin,rowlinson2013molecular} which, in turn, enables us to model electrowetting phenomena.~\cite{marinescu2010electrowetting,monroe2007distinctive}

The deformation of porous electrode during charging was proved by MD simulations. \cite{ori2014ionic, xu2020effects} It was shown \cite{ori2014ionic} that the equilibrium pore width slightly differs from the initial one, because of the layering inside the pore. Probably, the observed small deformations are due to the considered pore wall material, i.e., two thick silica slabs. The consideration of mechanical deformation within more flexible structures such as porous graphene,\cite{mendez2019performance} and, especially, in the carbon chains structures \cite{vatamanu2013increasing} is expected to dramatically change charging performance. Actually, layered electrodes consisting of a number of graphene or MXene sheets expand during charging up to $\pm15\%$.\cite{xu2020effects} Thus, the deformation of electrode is determined by electrode structure, material along with the ion layering and orientation. Despite these observations, it still is not clear how mechanical properties affect charging performance. We consider this issue to require a detailed analysis by theoretical and numerical approaches. 

In the ultratight pores, adsorption-induced deformation reaches the order of GPa and appropriate ion layering can cause both porous volume expansion and reduction. \cite{balzer2016deformation, gurina2022disjoining} Besides, the ultratight pores are known to create a superionic state for the confined ionic liquid and provide ionophobic conditions that are optimal for energy storage.\cite{vatamanu2013increasing, kondrat2016pressing, kondrat2023theory} Consequently, the interplay of confined effects together with the mechanical deformation of porous space is a key to answer whether it is attainable to create ionophobic conditions that way. Additionally, how mechanical deformation will change IL composition inside pores and modify the oscillating behavior of integral capacitance with pore width related to interfering EDLs? \cite{feng2011supercapacitor} Moreover, will the effects of mechanical deformation strengthen or weaken on geometrically rough surfaces? All these issues raise the lack of important fundamental insights, which need to be obtained in the future. The effect of surface roughness on mechanical properties in the confinement conditions was theoretically studied only for simple liquids.\cite{kolesnikov2020adsorption}

Regarding research on ionic liquid thin films, the main focuses here are on the film stability and search for optimal conditions for lubrication.\cite{mendonca2013nonequilibrium, smith2013quantized, david2017electrotunable, di2019electrotunable} To investigate the stability of ionic liquid films, researchers reproduce the squeezing out phenomena, i.e., the resistance of the film layers to the particular normal load pressures. In that way, the conditions of the film existence, the film structure together with structural forces, are obtained to reproduce and predict the results of SFM experiments. \cite{smith2013quantized, di2019electrotunable} It is also a preparation part for the investigation of electrotunable friction, where the friction properties of ionic liquid are controlled by the charges on the confining surfaces and the load pressure between them. \cite{mendonca2013nonequilibrium, smith2013quantized, david2017electrotunable, di2019electrotunable} It is worth to note, that ionic liquid film showed nonmonotonic friction properties with a charge on the surface at high loads caused by rheological transition, that is related to ion reorientation and reorganization. \cite{di2019electrotunable} Building an analytical approach or theory to describe this phenomenon would move the field forward. Since electrotunable lubrication is a recent research area, only numerical and experimental studies were carried, while, it can be described by the theoretical approaches. Particularly, the squeezing out phenomena and structural forces can be described within mean-field approaches \cite{budkov2022modified} or density functional theory \cite{evans1987phase, corrente2023deformation} by calculating the main components of stress tensor and non-diagonal components of the stress tensor reflect friction properties in the ionic liquid films. 

The study of ionic liquid film behavior within the structured confined conditions is more sophisticated due to complex heterogeneous surface structures and structural defects of real surfaces. The AFM experiments usually aim to obtain the structure of surface geometry, having a non-ideal complex character. The behavior of IL thin film within complexly structured confinement has been studied only with MD simulations.\cite{mendonca2013nonequilibrium, david2017electrotunable} The obtained results are controversial so far, the friction showed both increase\cite{david2017electrotunable} and reduction.\cite{mendonca2013nonequilibrium} The increase in the friction coefficient with roughness was explained by lateral structuring inside the structural pattern of surfaces. We suggest that this can also be related to the particularly ordered periodic structure of the surface. Whereas, if the surface pattern has random character, it would probably give another result. It was shown that disorder character of ion distribution in the interfacial layer near rough surfaces leads to the reduction of the friction coefficient.\cite{mendonca2013nonequilibrium} Therefore, the structure of ionic liquid rules the friction properties. The expansion of the existing theoretical approaches describing mechanical effects within slit geometry to the non-ideal complex surface geometries is desired. This motivates to develop theoretical approaches on stress tensors calculation for surfaces with complex surface geometry to predict friction properties passing computationally expensive numerical calculations.

Therefore, a first-principles approach that allows us to calculate the stress tensor of inhomogeneous ionic liquids is highly relevant for practical applications, as they are much more convenient than experiments or numerical simulations. One possible method for analyzing microscopic stresses is the Irving--Kirkwood method.~\cite{shi2023perspective, rusanov2001condition,rusanov2001three} This method involves averaging the Irving--Kirkwood microscopic stress tensor over the different microstates of a system, which is a difficult task in itself.~\cite{shi2023perspective} However, it can be done within the context of molecular dynamics (MD) simulations.~\cite{shi2023perspective,gurina2022disjoining,gurina2024disjoining} An alternative to Irving--Kirkwood technique has been formulated recently in series of papers.~\cite{budkov2022modified,brandyshev2023noether,budkov2023macroscopic,vasileva2023theory,budkov2023dielectric,budkov2023variational,brandyshev2023statistical} This methodology involves deriving the stress tensor from the grand thermodynamic potential. This is done by viewing the latter as a functional of suitable order parameters and application of the Noether's theorems.~\cite{hermann2022noether,noether1971invariant} It is called a thermomechanical approach (TMA),~\cite{budkov2024surface} since this approach self-consistently reunites the concepts of the grand thermodynamic potential and the stress tensor. Let us briefly discuss the major results obtained by TMA.

Recently, Budkov and Kolesnikov~\cite{budkov2022modified} applied thr Noether's first theorem to the grand thermodynamic potential of an ionic liquid as a functional of the electrostatic potential, establishing the local mechanical equilibrium condition in terms of the symmetric stress tensor. The obtained stress tensor consists of two components: the Maxwell electrostatic stress tensor and the hydrostatic isotropic stress tensor. The authors extended the local mechanical equilibrium condition to cases where external potential forces act on the ions. They then derived a general analytical expression for the electrostatic disjoining pressure of an ionic liquid confined in a charged nanopore slit, which extended the well-known DLVO expression to different reference models of ionic liquids. Brandyshev and Budkov ~\cite{brandyshev2023noether} propose a general covariant approach based on Noether's second theorem, allowing them to derive the symmetric stress tensor from a grand thermodynamic potential for an arbitrary model of inhomogeneous liquid. Note that this method is
similar to the one used by Hilbert in the general theory of relativity to derive the energy-momentum tensor from the action functional.~\cite{earman1978einstein} The authors applied their approach to several models of inhomogeneous ionic liquids that consider electrostatic correlations of ions or short-range correlations related to packing effects. Specifically, they derived analytical expressions for the symmetric stress tensors of the Cahn--Hilliard--like model,~\cite{maggs2016general} Bazant--Storey--Kornyshev model,~\cite{bazant2011double} and Maggs--Podgornik--Blossey model.~\cite{maggs2016general} Based on the variational field theory framework, the authors~\cite{budkov2023variational} extended the previous mean-field formalism~\cite{budkov2022modified} within the variational field theory of ionic fluids\cite{wang2010fluctuation_,lue2006variational_}, taking into account the electrostatic correlations of the ions. They employed a general covariant approach formulated in~\cite{brandyshev2023noether} and derived a total stress tensor that considers the electrostatic correlations of ions. This was achieved through an additional term that depends on the autocorrelation tensor function of local electric field fluctuations in the vicinity of the mean-field configuration. By utilizing the derived total stress tensor and applying the mechanical equilibrium condition, the authors established a general expression for the disjoining pressure of the ionic liquids confined in a pore with a slit-like geometry. To date, the TMA has been used to model the behavior of solvation pressure in nanopores containing liquid-phase electrolytes~\cite{kolesnikov2022electrosorption,vasileva2023theory} and polyelectrolytes,~\cite{budkov2023macroscopic,budkov2023dielectric} as well as to describe the surface tension of electrolyte solutions at the interface with air and dodecane.~\cite{budkov2024surface} 

Despite significant advancements in theoretical and numerical approaches for characterizing the mechanical properties of solid materials, numerous intriguing issues remain to be explored. For instance, the effect of electrode deformation on charging performance, particularly in relation to varying electrode structures and materials, requires further investigation. Furthermore, it is necessary to develop a theoretical framework to describe the non-monotonic friction behavior with respect to charge and normal forces in slit geometries, and to clarify the contradictory results observed in friction between structured surfaces. We hope to encourage further research in this important area.

\section{Chemical structure of material surface}
\label{op}

\begin{figure*}
\centering
\includegraphics[width=1\linewidth]{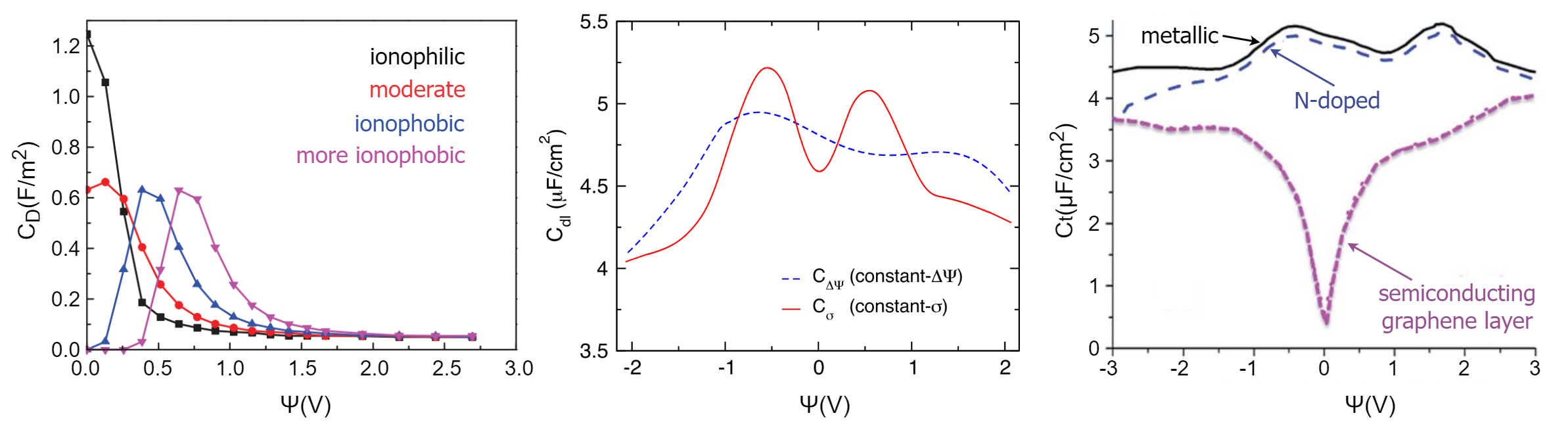}
\caption{The contribution of surface material chemical properties on differential capacitance (DC). On the left, DC profiles variation with surface ionophobicity are reproduced from Ref.\cite{lian2016can}. In the center, the effect of surface polarization on the DC profile is reprinted from \cite{haskins2016evaluation}, with the permission of AIP Publishing. On the right, the impact of the chemical structure of confining surface on the DC profile is taken from Ref.\cite{vatamanu2017charge}}
\label{fig:chemistry}
\end{figure*}

Generally, ionic liquid can be confined between surfaces made of various materials depending on the application. For supercapacitors based on EDL, called EDLC, electrodes commonly made of various nanostructured carbon materials, such as activated carbons (ACs), carbon nanotubes (CNT), carbon aerogels and others, \cite{simon2008materials, liu2010advanced} but recently metal-organic framework (MOF) electrodes have also attracted considerable attention. \cite{feng2018robust, bi2020molecular} A tribological research, such as surface force balance (SFB) experiments and atomic force microscopy (AFM), can be carried out with any material of different nature, like mica sheets or slabs,\cite{smith2013quantized, david2017electrotunable} metallic slabs,\cite{mendonca2013nonequilibrium} or silica colloid on gold surfaces \cite{sweeney2012control} and others. \cite{werzer2012ionic, mendoncca2013novel} Actually, AFM cantilever is usually compounded from silica or silicone nitride \cite{bhushan2008afm, raman2008cantilever, onofrio2013molecular} or can also be made of diamond, \cite{gao2007atomic} and gold. \cite{pishkenari2010surface} The chemical structure of electrode material determines both the intermolecular (non-electrostatic, van der Waals) interactions and the electrostatic interactions between ions and electrode wall. The former can be characterized by selectivity and wettability, or considering charged systems - ionophobicity. The second is determined by polarizability, i.e., the ability of the electrode to respond to the ionic liquid in contact. 

The influence of electrode material chemical structure on the behavior of ionic liquid in the confinement was considered by several MD studies. \cite{lahrar2021simulations, li2013dynamic, singh2013molecular, bacon2023key} Generally, when electrode-electrolyte system is considered the role of non-electrostatic interactions is neglected. However, it was shown by MD simulation that surface material can change the IL structure. \cite{bou2023effect} For the confined system these forces become more significant and define IL behavior, for instance, ion diffusion. \cite{li2013dynamic} In the MD study \cite{singh2013molecular} the authors performed the comparison of the pore wall material (rutile or graphitic) impact on the structure and dynamic of ionic liquid in the slit pore. As a result, the strength of the interactions between the pore walls and the ions can significantly decrease ion diffusion and change the structure of the confined IL, i.e., IL density and ion orientation. Therefore, carbon materials are effective for charge storage applications also due to low intermolecular interactions with IL ions. In another MD study, \cite{bacon2023key} the authors showed how ion-electrode attraction slows down ion exchange during charging, leading to lower capacitance, and modifies the IL structure and ions orientation. Besides, it was observed that the presence of functional groups with different chemical structures (ester, hydroxyl, anhydride acid, and carboxyl functional groups) on the electrode surface of ZTC can improve ion diffusion and decrease ion density near the surface, keeping IL structure. \cite{lahrar2021simulations} Thus, we can conclude that the chemical structure of electrode or inclusion of electrode surface modifiers is one of the main parameters for optimal charging performance.

Regarding ionophobicity, this characteristic is usually described as classification: ionophilic or ionophobic, i.e., filled with ions or empty uncharged pore. It is well known that ionophobic pores provide higher energy density and faster dynamics, so it is one more key factor in creating effective energy storage tools.\cite{kondrat2023theory} The possible ways to create ionophobic pores are to consider very tight pore \cite{mo2020ion} or structured electrode,\cite{vatamanu2013increasing} to add solvent, \cite{rochester2016charging} to select a particular electrolyte, \cite{lian2018electrochemical} and electrode surface materials or coating.

While we shouldn't aim to create absolutely ionophobic pores for optimal charging performance. Perhaps, a small modification of electrode material or surface structure leading to more ionophobic conditions will be enough. So we appeal to treat this characteristic more accurately, similar to wetting phenomena, like in the following works,\cite{vella2016capacitance, lian2016can} by tuning the appropriate chemical composition of electrolyte and electrode, rather than in a typed way. 

Naturally, atoms of the electrode surface adjust to the ionic liquid to maintain equipotential conditions by electrostatic interactions. The electrode can be considered as metal or ideal polarizable, atomic polarizable (non-ideal), or non-polarizable at all. The electrode polarizability explains the formation of the superionic state in the confinement, so it is crucial to take this characteristic into account in the confined systems. \cite{kondrat2010superionic, kondrat2023theory}
Polarizability affects dynamical properties, such as diffusion or time scale of polarization relaxation or relaxation phenomena, \cite{merlet2013simulating, zeng2021modeling, gading2022impact} differential capacitance, \cite{haskins2016evaluation} local ion structure near electrode surface at high voltages,\cite{merlet2013simulating} and heating rate. \cite{merlet2013simulating, zeng2021modeling} Insignificant effects were also observed for local ion structure and diffusion coefficient at low voltages, which is explained by the leadig role of steric interactions rather than electrostatic one. \cite{merlet2012molecular, ntim2020role, breitsprecher2015electrode, haskins2016evaluation, gading2022impact} No effect was observed on friction properties that can be explained by strong screening or lack of structural modifications.\cite{di2021structural}
Electrode polarizability is determined by boundary conditions in analytics and theory \cite{bruch2022thermodynamics} or by the application of constant potential (CPM) or constant charge methods (CCM) in simulations. 

Briefly, CCM methods are computationally easier to implement but provoke unphysical results like too fast dynamics \cite{merlet2013simulating, zeng2021modeling, gading2022impact} and extreme energy dissipation. \cite{merlet2013simulating, zeng2021modeling} Therefore, it is better to implement CPM to represent proper physical phenomena. 

One more effect, related to the material of the electrode, is quantum capacitance. \cite{luryi1988quantum, kondrat2023theory} It occurs in the carbon materials, especially low-dimensional such as CNT or graphene, \cite{pomerantseva2019energy} which are of the main interest for energy storage applications. \cite{verkholyak2022less, kondrat2023theory} Quantum capacitance reflects the ability of an electrode to accumulate charge in response to potential on the electrode. \cite{verkholyak2022less} For metal electrode quantum capacitance is infinite, but for non-metal materials it is finite and related to the limitations of density of states for electrons in the electrode material. \cite{verkholyak2022less, kondrat2023theory} 
The limitation of density of states for graphite electrodes explained lower capacitance in comparison to metal electrodes. \cite{gerischer1985interpretation}
For CNT the density of states can be calculated analytically, \cite{mintmire1998universal} while for the other materials, one should carry quantum-mechanical calculations. \cite{paek2012computational} Interesting, that quantum capacitance can be modified with mechanical deformation.\cite{paek2012computational} While the impact of quantum capacitance is still under debate, \cite{verkholyak2022less, kondrat2023theory} no doubt, that it should be considered to provide an accurate description of charge storage properties.

\section{Challenges for further research}

In summary, despite advancing research that takes into account the properties of confining materials, several unresolved issues remain, particularly regarding ion dynamics in slit geometries. For instance, to clarify the discrepancies on the DC behavior with charging mechanisms as well as modifications of ion dynamics with pore width. Afterwards, the study of ion dynamics in rough confinement condition and nanoporous systems should be continued.

Moving on, for the investigations of surface geometry influence, it is essential to extend both numerical and theoretical methodologies. Researchers should focus on uncovering the fundamental mechanisms behind observed phenomena while aiming to bridge the gap between theory and simulation. We should definitely know how the form, size and number of structural features will impact the EDL and how to create more optimal EDL structure for efficient charge storage. Along with that we should consider more complex physical behavior of the system including ion correlations, ion pair formation, and other relevant effects. 

Subsequently, the analysis can be further complicated by the consideration of confinement conditions accounting for the specificity of confined systems and distinctly separating the underlying reasons of observations. Important to note, that in the confinement, reliance on more rigorous theoretical frameworks, such as density functional theory (DFT), is necessary to accurately represent phenomena at the molecular scale.

Another critical issue is the consideration of disordered nanoporous structures and the integration of processes occurring within these complex geometries. The upscaling problem is longstanding and extends beyond electrochemistry to various fields, including oil recovery and others. Therefore, it is essential to compile existing efforts aimed at addressing this problem and to review recent advancements across each relevant discipline.

Regarding mechanical properties of confining materials there are enough theoretical approaches that can be applied to the investigation of ionic liquids in relation to charging and friction phenomena. It is essential to understand the role mechanics in charging process and to optimize the investigations of friction in the ionic liquids films. Additionally, it is valuable to build approaches that integrate mechanics with complex surface geometries, as this could provide new insights into the behavior of ionic liquids under confinement.

The numerical approaches are the main tool for advancing our understanding of molecular-scale systems, Therefore, it is necessary to continue investigating confined ionic liquids especially concerning ion dynamics and friction. Besides, efforts should be directed toward description of more complex physical phenomena with aiming on the connection of the simulations and theory.

In the following sections, we present a comprehensive discussion of potential extensions to current methodologies aimed at addressing the mentioned issues.

\subsection{Theoretical models}
\label{TM}

Analytical continuum models hold significant practical value for understanding the intricate dynamics of electrolytes, with a particular emphasis on ionic liquids, within confined pores comprising multiple fluid layers. Balancing predictive accuracy with computational efficiency, a notable category of models are those of the Ginzburg--Landau type, yielding nonlinear differential equations. These models, exemplified by modified Poisson--Fermi equations,\cite{kornyshev2007double, bazant2011double, goodwin2017underscreening} offer a promising avenue for analysis. In certain scenarios, these equations permit analytical solutions, particularly evident under conditions of low applied electrostatic potentials. Such analytical frameworks provide valuable insights into the behavior of electrolytes in complex confined environments, offering a nuanced understanding of their interactions and properties.

However, such models were initially developed with the assumption of ideally smooth pore walls, and taking into account roughness, patterning and corrugated surface morphology in such models represents a pressing problem, despite recent progress in this direction.\cite{aslyamov2021electrolyte, shen2022effect, khlyupin2023molecular} Particular challenges arise when we talk about rough surfaces on the molecular scale, which begin to play a significant role in nanoconfinement. The complexity arises from two independent ways in which surface heterogeneity at this scale influences the physics of the electrolyte:

\begin{enumerate}
\item First, the equipotential conditions for a conductive material require defining the boundary conditions in the electrostatic potential equation (Poisson--Boltzmann for example) on a rough surface. Due to the long-range Coulomb interaction, such a disturbance at the boundary should be transmitted further away from the wall and can affect the structure and macroscopic properties of the electrolyte, like capacitance or charging dynamics.
\item Secondly, when the roughness scale approaches the characteristic size of electrolyte molecules, it is necessary to take into account the asymmetry of ion sizes, since now these two characteristic spatial scales (the scale of roughness and the characteristic sizes of ions) are of the same order of magnitude and cannot be neglected. This is especially true for ionic liquids, where positively and negatively charged ions can differ in size by several times.\cite{vatamanu2011influence, xing2012nanopatterning} Differences in size and shape of ions near a rough surface on the molecular scale led to ion-specific soft repulsion potentials due to different depths of ion penetration into the surface \cite{khlyupin2017random, khlyupin2023molecular}. These potentials must be taken into account in free energy models of the electrolyte, which will also affect its structural, capacitance and dynamic properties.
\end{enumerate}

We emphasize that both contributions are related to the characteristic molecular scale of surface heterogeneity and cannot be omitted without strong arguments. In pioneering works, Daikhin, Kornyshev and Urbakh \cite{daikhin1996double, daikhin1998nonlinear} investigated the problem of the influence of boundary conditions specified on curved random surfaces on the capacitance properties of the electric double layer. They investigated both linear and nonlinear models based on classical Poisson--Boltz-\\mann theory at the metal-electrolyte/semiconductor interface.

The authors showed that the double layer capacitance is highly sensitive to the interplay between the Debye length and the roughness parameters: the root mean square height of roughness and lateral correlation length. It is interesting to note that at voltages close to the potential of zero charge, the main effect is caused not by fluctuations in the surface height, but by the lateral correlation length. Thus, it became clear that one parameter describing the surface roughness only in terms of the root mean square height is not enough to build a correct EDL model.

Following the authors, \cite{daikhin1996double, daikhin1998nonlinear} we note the important limitations of their model: the weakly rough surfaces model for which lateral disturbances have a larger spatial scale than fluctuations along the normal is considered; A model with a symmetrical electrolyte with ions of the same size and charge was studied. Thus, this approach made it possible to take a step forward regarding the first problem of roughness, which was discussed above. However, at the same time, it does not take into account ion-specific repulsion near the wall and is suitable for describing large-scale smooth curved surfaces far from the molecular scale. It is also important to note that the model is based on the classical Gouy--Chapman theory with the Poisson--Boltzmann equation for the electrostatic potential, containing only a second-order derivative and does not take into account more complex effects associated, for example, with ion correlations.

On the other hand, recently progress has been made in theoretical models that study the interaction of asymmetric electrolytes with roughness on a molecular scale, taking into account ion-receptive potentials. \cite{aslyamov2021electrolyte, khlyupin2023molecular} Namely, it was shown that molecular scale effects including ion size asymmetry and electrode surface roughness significantly contribute to ions separation in EDL.\cite{khlyupin2023molecular}  Authors proposed novel analytical models that provide PZC and DC solutions for EDL in contact with the rough electrode surface. Using this model, it was possible to describe a number of phenomena observed in the recent numerical and experimental studies: formation of the third peak in a DC profile by structural (reorientation) transition of IL ions; DC shape transition from bell to camel and backward due to the roughness of an electrode surface; nonmonotonic behavior of DC on rough surface with temperature and ion size ratio. However, in the considered model, the boundary conditions on a rough wall are taken into account in a simplified manner and it is assumed that the main effect is achieved due to ion-specific steric interactions with the electrode surface.

To summarize, separate theoretical studies are devoted to two mechanisms of the influence of roughness on the properties of EDL. From our point of view, it is of great interest to combine these approaches within the framework of an analytical model, which will be based on a single model of a rough surface with a single set of parameters describing the roughness. Then, within the framework of such a model, it will become possible to strictly analytically compare the contributions of the two mechanisms of influence of roughness and determine the regions of parameters when the first or second mode significantly predominates. In addition, based on molecular modeling data, it seems that steric rather than electrostatic interactions determine the local structure at typically high ionic densities.\cite{kondrat2023theory} Confirming or disproving the small contribution of electrostatics to the EDL properties in such systems is of great interest, since this will significantly simplify the development of more complex models (such as DFT based).


It is reasonable to take the first steps of accounting molecular surface roughness on the basis of simple and physically transparent model (like Poisson--Fermi type model), because it allow us to build the analytical solution and clearly present the changes in all important physical quantities in the system. However, it is of great interest for further research to use more advanced models and obtain new effects of molecular scale roughness on the structure of EDL for systems with more complex physics like strong Coulomb ion-ion correlations. Room Temperature Ionic Liquids (RTILs) exhibit robust short-range Coulomb correlations among ions, coupled with hard sphere repulsion arising from finite ion sizes. This interplay leads to a transition from overscreening phenomena, characterized by charge density oscillations, to the crowding regime, where dense layers of counterions accumulate at the interface. \cite{bazant2011double, de2020interfacial} Importantly, each regime manifests its own capacitance-voltage relationship. To elucidate the overscreening and crowding dynamics in RTILs, Bazant, Storey, and Kornyshev introduced a theoretical framework \cite{bazant2011double}. In their work, they propose a Ginzburg--Landau type functional for the total free energy, incorporating a higher-order gradient term in the electrostatic potential to account for overscreening in strongly correlated liquids. This term accounts for the correlation length of electrostatic interactions. By minimizing the free energy functional with respect to the potential field and incorporating crowding effects using a lattice-gas approach, one can derive a modified Poisson-Fermi equation featuring a fourth-order derivative term. This term governs deviations from mean-field theory. 

Exploring this equation in conjunction with rough surfaces could yield valuable insights. It would be particularly promising to investigate the interplay between surface roughness parameters and applied potentials to identify conditions under which overscreening is disrupted in the electric double layers. Since in the BSK model another characteristic scale parameter (correlation length) is added to the usual Debye length, we should expect an interesting picture of physical regimes due to the interplay of these model parameters and the parameters of surface roughness. It is also worth paying attention that the BSK equation contains an additional quadruple derivative of electrostatic potential and the analysis of the influence of boundary conditions on a rough surface for such model is still an open question that needs to be solved. In a recently proposed model,\cite{nesterova2024mechanism} the authors demonstrate that surface roughness can either enhance or diminish the overscreening effect, conditional to the relationship between the correlation length and the scale of surface structural inhomogeneities. Notably, the model predicts a reduction in differential capacitance due to the soft repulsion of larger ions from the electrode surface. However, this initial exploration necessitates further development, as the presented model includes considerable assumption related to boundary conditions that require refinement. Thus, studying the conditions for overscreening breakdown near the surfaces with complex structure poses a challenge to the theory. Moreover, if such a BSK type model is expanded to the confinement conditions, then it will become possible to describe the change in ion composition in nanopores like in MD \cite{vatamanu2015non} or to investigate for the effect of pore wall structure on anomalous capacitance.


Another possible physical extension of the theory is to account for more complex ion behavior, particularly to consider ion bonding near rough electrode surface. Experimental observations of extremely long screening lengths in RTILs suggest a tendency for ions to aggregate into neutral clusters.\cite{gebbie2015long} This assumption was proved by molecular dynamics simulations and theoretical analysis of velocity-autocorrelation functions,\cite{feng2019free} revealing that only approximately 15-25\% of ions exist in a free state. Later, Goodwin et al. \cite{goodwin2022gelation} tackled the Electric Double Layer (EDL) conundrum by comprehensively addressing all clusters formed through associative bonds, including the formation of ion networks or gels. However, this model necessitates numerical solutions. Subsequently, they proposed an analytical formulation for this model, limiting formations to ion pairs. \cite{goodwin2022cracking}

We anticipate that surface roughness will significantly influence cluster formation and equilibrium distribution in ionic liquids. By incorporating soft ion-specific repulsion potentials, influenced by surface roughness, into the cluster model, \cite{goodwin2022gelation} one should consider the dependence of such potentials on various cluster sizes. Thus, surface roughness is expected to impose constraints on the size of clusters that can form at a specific distance from the electrode. In essence, surface irregularities prevent larger clusters from occupying certain areas. Consequently, clusters of different sizes will have distinct penetration depths, with maximum depth for free ions and decreasing depths as cluster size increases. Gels, being clusters of infinite size, can only form away from the surface. Consequently, the varying penetration depths for each cluster size result in shifted concentration distributions. This model warrants detailed examination and numerical analysis. At this stage, we can only posit hypotheses and expectations.


Poisson-Fermi type models are sufficiently straightforward to be applied to a wide range of problems. However, these models have significant limitations; specifically, their local character precludes an accurate representation of electrostatic correlations at the molecular scale, thereby hindering their ability to effectively capture screening behavior. To address the limitations highlighted in the study by de Souza et al., \cite{de2020interfacial} the authors take a significant step forward in refining the ionic liquid (IL) model by enhancing the representation of crowding effects. They introduce an additional term in the free-energy functional, known as the weighted densities approximation (WDA), which accounts for the volumetrically weighted densities of local ion concentrations. In their approach, they employ the Carnahan--Starling equation of state as a reference for a hard sphere system. Consequently, this density functional theory (DFT) based model is proficient in predicting discrete layering and extended overscreening with a longer screening length. However, the resulting model is represented by a system of integro-differential equations and necessitates numerical solutions.

To take into account roughness in EDL, one can adopt DFT based models originally developed to characterize Lenard-Jones (LJ) confined fluids in nanopores, incorporating molecular scale surface roughness. These models demonstrate the disruption of layering structures in the presence of roughness compared to an ideal smooth pore surface, a phenomenon supported by experiments on low-temperature adsorption. For instance, in the works \cite{aslyamov2017density, khlyupin2017random} authors introduce a novel random surface density functional theory (RSDFT) formulation tailored for geometrically heterogeneous solid surfaces, crucial for describing the thermodynamic properties of confined fluids. A distinguishing feature of this theoretical approach is its accurate representation of solid surface geometry, considering correlation properties in both lateral and normal directions. Comparisons with experiments on gas adsorption validate the efficacy of this method. In another works \cite{aslyamov2019random, aslyamov2019theoretical} authors develop a new version of the SAFT-DFT approach named RS-SAFT, capable of describing the adsorption of chain molecules on geometrically rough surfaces. This innovative approach combines RSDFT with Statistical Associating Fluid Theory (SAFT). \cite{chapman1989saft}

Consequently, there is a significant interest in applying advanced models developed for confined LJ fluids to IL systems with molecular scale surface roughness. It is important to emphasize that what was said earlier about the influence of the parameters of the geometry of a rough surface on the properties of the electrolyte remains true also in relation to DFT models. Therefore, for the accurate development of DFT for rough surfaces, it is necessary that the surface model contains both the fluctuation parameters in height and the correlation length in the lateral direction (see the recent review \cite{kazemi2024wettability} for a comparison of the used rough surface models within the framework of existing DFT's).


In the realm of theoretical development, there's a notable focus on low-dimensional models, particularly in the 1D and 2D domains. These models serve as valuable tools for understanding phenomena of superionic state within conductive strong nanoconfinement, where electrostatic interactions are effectively screened. \cite{kondrat2023theory} In a recent study by Kondrat et al., \cite{kondrat2010superionic} a phase transition from an ordered state at low potentials to a completely disordered phase was observed in 2D slit-shaped pores using the Blume--Capel spin model. This transition occurs as the voltage increases, reaching a point where the ordered structure is disrupted. An intriguing finding was the peak in capacitance at the transition, indicating a shift in the region of "active charging."

Traditionally, these models have only considered perfectly smooth walls. However, there's growing interest in exploring how wall curvature might influence predicted phase transitions or even give rise to new phases, such as an "ionic glass." Moving forward, the extension may be inspired by incorporating the interionic potential for the superionic state within rough pore wall geometries. This could involve applying spin glass models with quenched disorder, considering the additional random correction to pair interaction potentials between ions. Similar phenomena have been well-studied in spin glasses with random frozen disorder, such as the Sherrington–Kirkpatrick model, \cite{sherrington1975solvable} which could lead to an additional spin glass transition under certain parameters. It's worth noting that similar investigations of the effects of smooth random surfaces on classical LJ fluid monolayers have been conducted, \cite{khlyupin2016effects} revealing shifts in phase transition temperatures due to quenched random interactions and random site fields. This interdisciplinary approach holds promise for further elucidating the complex behavior of confined ionic systems and may pave the way for novel applications.


Regarding mechanical problems, despite the fact that the TMA has been used to describe mechanical stresses in liquid-phase electrolytes, there are several issues that need to be addressed in the near future when it comes to ionic liquids under nanoconfinement. First, the TMA has been applied to obtain stress tensors within the framework of mean-field models of ionic fluids, which are based on 'local' thermodynamic potentials such as free energy functionals containing spatial derivatives of order parameters (potentials and concentrations) up to second order.~\cite{brandyshev2023noether} However, it is still unclear how this approach can be extended to more sophisticated theories based on the free energy functionals that depend on weighted order parameters,~\cite{de2020interfacial,de2022structural,de2022polar} such as the case of hard sphere models.~\cite{roth2010fundamental} However, progress can be made by considering the case where the weighting function is the Green's function of a certain differential operator.\cite{budkov2024} Furthermore, it was possible to calculate the stress tensor for essentially non-local functionals that describe the fluctuation correction to the mean-field thermodynamic potential within statistical field theory.~\cite{budkov2023variational,brandyshev2023statistical} Second, the TMA allows us to calculate mechanical stresses not only in ionic liquids confined in charged pores with a slit geometry (where we can use the contact value theorem~\cite{dean2003field,buyukdagli2023impact}), but also in pores of arbitrary geometry, including those with non-convex geometries and corrugated walls. Additionally, it will be grateful to build the analytical model, basing on the TMA approach, which can describe the observed nonmonotonic friction behaviour related to structural transition phenomena of the ionic liquid films under charged confinement conditions.~\cite{di2019electrotunable}

\subsection{Molecular Dynamics Simulations}
\label{MD}

The challenging task for MD simulations is a correct description of interactions in ionic liquids,~\cite{parker2018molecular,bedrov2019molecular} as well as their interaction with surface. In the molecular modeling, the common approaches are electron Density Functional Theory (eDFT), polarizable all-atom force fields, all-atom force fields with fixed charges, united-atom and coarse-grained approaches. The eDFT methods are the most precise for the computation of interatomic forces. However, they are computationally expensive for the simulations of system sizes, that are required for this area of research. Most of all, eDFT is used for the development of force fields. For example, Dommert and Holm~\cite{dommert2013refining} use the \textit{ab initio} calculations of bulk systems for deriving partial charges for the classical force field. Akkermans et al. extend the classical force field for the ionic liquids based on the eDFT calculations.~\cite{akkermans2021compass}

The first examples of polarizable force fields for ionic liquids could be found in the works by Voth et. al~\cite{del2004structure} and Borodin.~\cite{borodin2009polarizable} The works show that the dynamic calculation of atomic polarisabilities allows to enhance the prediction of dynamic properties of ionic liquids, such as the ion self-diffusion coefficients, conductivity and viscosity. The complete review of these approaches (Drude oscillators and induced point dipoles) could be found in the work by Bedrov et al.~\cite{bedrov2019molecular} The authors conclude that polarizable force fields should be used for calculations of dynamical properties, while the non-polarizable force fields reproduce accurately the local structure and static properties of ionic liquids. The CL\&Pol polarizable force field is specially developed for the simulations of ionic liquids by Padua's group,~\cite{goloviznina2019transferable,goloviznina2021extension} repository for CL\&Pol is available.~\cite{pad_git} The important fact is that the polarizable force fields are up to 10 times more computationally expensive.

The force fields with fixed charges are applied by Maginn's group~\cite{morrow2002molecular,maginn2009molecular,zhang2015molecular,Zhang2015} and other collectives.~\cite{margulis2002computer,de2002computational,liu2004refined}
The guidelines for the simulations with fixed charges are published recently.~\cite{Sun2022} The link~\cite{il_parameters} contains the parameterization files for the ionic liquids. The examples of united-atom and coarse-grained force fields for ionic liquids are the works by Ludwig et al.~\cite{koddermann2013comparison} and Laaksonen et al.~\cite{wang2013multiscale} The highly computationally efficient coarse-grained models are shown to reproduce nanostructural organization and liquid–liquid extraction for imidazolium-based ionic liquids.~\cite{vazquez2020martini}

The methods for the transport properties calculation of ionic liquids have developed during last decades. The diffusion and viscosity coefficients of liquids could be calculated from MD exactly (for details see~\cite{maginn2019best,kondratyuk_ufn}). The methods for the diffusion are: Einstein-Smoluchowski and Green-Kubo. For example, Feng et al. use the Green-Kubo relation for the calculation of diffusion in [BMIM][TFSI].~\cite{feng2019free} The MD methods have proven that the diffusion of ions in solutions depend significantly on the properties of the solvation shells of ions.~\cite{1,2,3} The diffusion and the occurrence of various electrochemical reactions are connected in solutions.~\cite{4,5} The Maginn's group applied the equilibrium and non-equilibrium MD simulations for the calculation of viscosities of ionic liquids.~\cite{zhang2015molecular} They developed the time decomposition method~\cite{Zhang2015} for the convergence of the Green--Kubo integral. At the moment, there are not many studies of ion diffusion inside rough pore walls.

The transport properties of ionic liquids could change significantly in the confinement. The current challenges of studying rough geometry in MD are the lack of data on the interaction parameters for surface atoms and ionic liquid. Only a few surface types are available at the moment, so the further investigation is required. The another challenge with MD simulations of surfaces with random roughness is the amount of sampling for the statistical relevance of results. The atomic scale rough surfaces could be generated using various approaches. For example, PoreMS python package~\cite{kraus2021porems} could generate rough pores of different sizes. Also, one could find Pyrough,~\cite{iteney2024pyrough} which is a tool for building 3D samples with rough surfaces for atomistic simulations. The useful tool for the analysis of geometric properties of porous media is PoreBlazer.~\cite{sarkisov2020materials}

The electrotunable friction and usage of ionic liquids as lubricants are promising areas of research. The main characteristics for this applications are the level of friction and shear viscosity dependencies on various parameters. The examples are the distance between electrode walls, the roughness of walls and the thermodynamic parameters. The viscosity of confined ionic liquid could also be obtained from the molecular dynamics simulations. For the confined systems, the non-equilibrium methods are used mostly. They are based on the creation of the shear stress~\cite{Evans1990,Tuckerman1997,Davis2006} or the momentum flux (M\"{u}ller--Plathe method~\cite{Plathe1999, Bordat2002}) in the unit cell, applied to the confined liquid. The viscosity characterizes the velocity profile of established flow. The ionic liquids are highly viscous liquids, so the problem with this family of methods is the non-Newtonian response at the achievable within MD strain rates. This problem is similar to the calculations of liquids viscosities at high pressures.~\cite{Jadhao2017,Jadhao2019,kondratyuk2019}

In the confinement, the equilibrium methods also could be applied, which estimate viscosity at ``zero'' share rate. These viscosity values are close to the experimental and industrial applications. The interpretable method is based on the dependence of diffusion coefficient on the size of computational unit cell.~\cite{Yeh2004,Naberukhin2018,Orekhov2021,deshchenya2022} From the calculations of diffusion coefficient at different unit cell sizes, one could estimate the viscosity of confined ionic liquid from the linear slope. The only problem with this method is that Yeh and Hummer~\cite{Yeh2004} developed the equation for the neutral particles, so the electrostatic interactions need to be included. To the best of our knowledge, this correction is not accounted at the moment. The another problem with the equilibrium methods, such as Green--Kubo approach, arises in the correct description of the virial. The work~\cite{huang2014green} contains the theory for the application of the Green--Kubo relation to the solid-liquid interfaces directly from the equilibrium MD.

Also, the interesting idea is the link between the transport coefficients and excess entropy of ionic liquids. The two-particle contribution is in many cases the dominant contribution to the excess entropy.~\cite{Dzugutov1996,gao2018} The pairwise entropy can be directly computed from the radial distribution function~\cite{khrapak2021,nicholson2021} even for polyatomic molecules.~\cite{nikitiuk2022pair} Malvaldi and Chiappe~\cite{malvaldi2010excess} have shown the scaling between excess entropy and diffusion coefficient of 1,3-dimethyl imidazolium chloride. The challenging question is the calculation of excess entropy of ionic liquid in confinement, which could indicate its influence on the local structure. For the estimations of excess entropy, the thermodynamic integration could be applied.~\cite{nikitiuk2022pair}

\section{Conclusion}
\label{conclusion}
	
In this review, we have underscored the promising potential of room temperature ionic liquids (ILs) as electrolytes in various technological applications, notably in energy storage and electrotunable lubrication for AFM experiments. Further investigation is required for their behavior description in the structured confinement conditions along with confining material properties. While previous studies have primarily concentrated on enhancing IL characteristics through electrolyte property selection, the impact of confining material properties in these systems have received limited attention.

Addressing this gap, we have provided recent developments pertaining to the impact of electrode properties on IL behavior within confinement. Our analysis reveals the critical role of the structural, mechanical, and chemical properties of the confining materials in ionic liquid behavior in the confinement. Although the presented findings highlight the significance of electrode properties, the proper characterization of confined ionic liquids remains challenging and further comprehensive research is still ahead.

An important aspect of this review was to propose the avenues for future investigations of the confined ionic liquids with accurate consideration of surface material properties. Particularly, the review reflects our perspective and vision concerning the prospective directions for theory and molecular modeling. We have dedicated attention to heterogeneous electrode surface morphology and the development of models for accurate analysis of roughness effects on electrolyte physics, particularly emphasizing key features and potential pitfalls along this trajectory. Furthermore, we have shared hypotheses and conjectures on addressing certain challenges.

Another notable observation from our literature analysis is the disparity between molecular dynamics (MD) methods and theoretical models. MD has significantly advanced compared to analytical models in describing key physical effects, offering immense potential for conducting numerical experiments in systems otherwise unobservable in real laboratories. These MD results inspire theoretical advancements and provide means to validate theoretical predictions, enriching the field as a whole. Moreover, there are contradictory observations in a number of MD studies that need to be resolved by fundamental theoretical vision.

Regarding the synergy of various approaches, we contend that many challenges concerning the influence of surface morphology on liquid system behavior at the molecular scale have been addressed in simpler Lennard-Jones systems. Considerable progress has been made in developing theory for such classical LJ systems in confinement with imperfect walls, suggesting the applicability and adoption of certain methods for charged systems like ILs.

In summary, our review not only underscores the importance of electrode properties in IL systems but also advocates for interdisciplinary collaboration and the adoption of methodologies from related fields to address existing challenges and propel research in this exciting domain.

\section*{Acknowledgements}

This research was supported by the Russian Science Foundation grant 24-67-00026. Authors would like to express sincere gratitude to our colleagues - A. Cherkasov, A. Ipatova, R. Sirazov for fruitful discussion that helped in shaping our thoughts as presented in this paper.

\section*{Conflict of Interest}

The authors declare the absence of any conflict of interest.

\begin{shaded}
\noindent\textsf{\textbf{Keywords:} \keywords} 
\end{shaded}


\setlength{\bibsep}{0.0cm}
\bibliographystyle{Wiley-chemistry}
\bibliography{refs}

\clearpage


\section*{Entry for the Table of Contents}



\noindent\rule{11cm}{2pt}
\begin{minipage}{5.5cm}
\includegraphics[width=5.5cm]{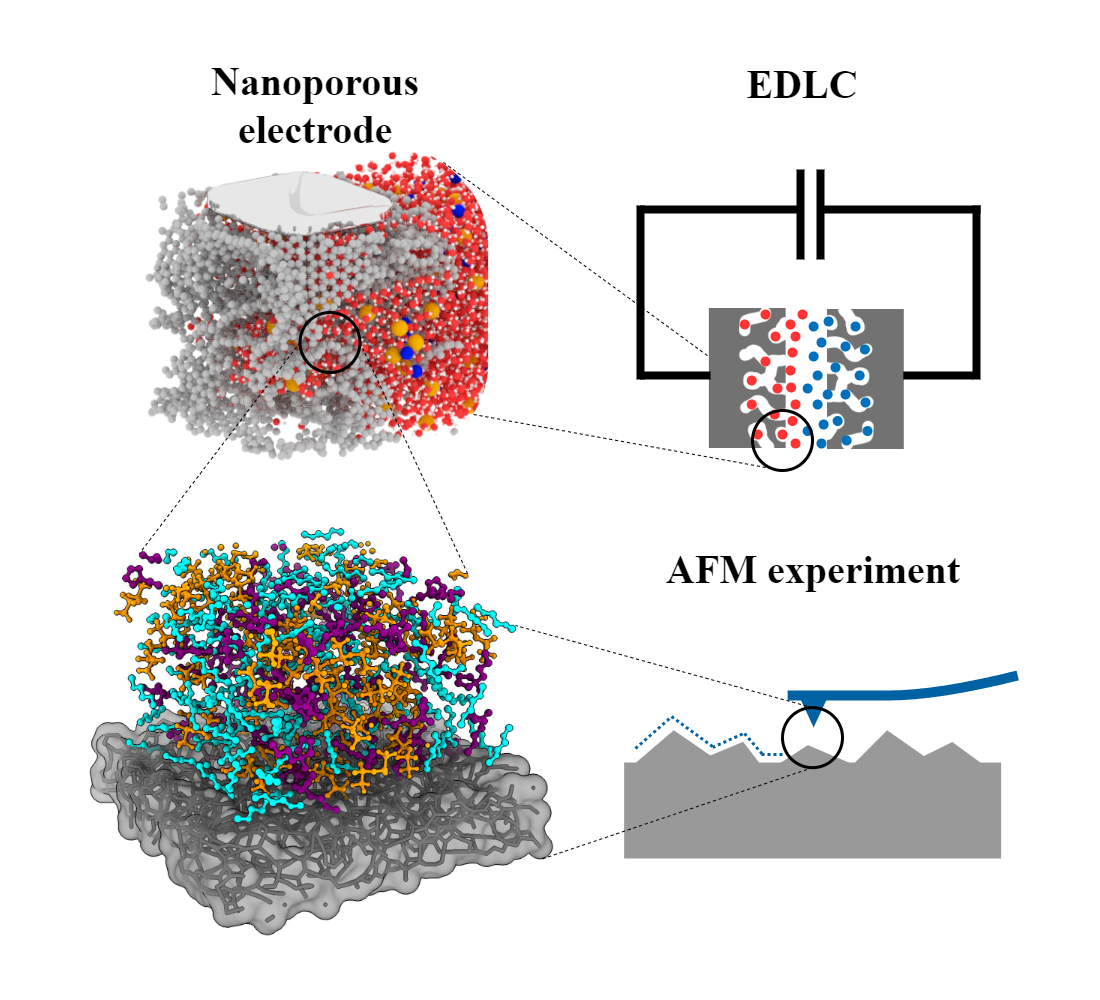} 
\end{minipage}
\begin{minipage}{5.5cm}
\large\textsf{
This review emphasizes the essential role of surface material properties in determining the behavior of confined ionic liquids. The analysis takes into account the significance of surface morphology and complex nanoporous structures, the mechanical and chemical characteristics of confining materials. In addition to extensive review of the literature, we also propose the avenues for further research.}
\end{minipage}
\noindent\rule{11cm}{2pt}

\vspace{2cm}


\begin{minipage}{11cm}
\includegraphics[width=11cm]{TOC.png} 
\end{minipage}




\end{document}